%% file: symtraces.tex
\documentclass[conference]{IEEEtran}
\usepackage{url}
\usepackage{amsmath,amssymb,amsthm}
\usepackage{stmaryrd}
\usepackage[vlined,figure,linesnumbered]{algorithm2e}
\usepackage{balance}
\usepackage{booktabs} 
\usepackage{url}
\usepackage{color}
\usepackage{listings}
\usepackage{cite}
\newcommand\lt[1]{{\lstinline+#1+}}
\usepackage{tikz}
\usetikzlibrary{calc}
\usetikzlibrary{shapes}

\newtheorem{definition}{Definition}

\renewcommand{\implies}{\Rightarrow}
\newcommand{\hide}[1]{}
\newcommand{\ignore}[1]{}

\newcommand{\tool}{\textsf{SymInfer}}
\begin{document}

\title{{\tool}: Inferring Program Invariants using Symbolic States}
\author{\IEEEauthorblockN{ThanhVu Nguyen}
\IEEEauthorblockA{
University of Nebraska-Lincoln, USA\\
tnguyen@cse.unl.edu}
\and
\IEEEauthorblockN{Matthew B. Dwyer}
\IEEEauthorblockA{University of Nebraska-Lincoln, USA\\
dwyer@cse.unl.edu}
\and
\IEEEauthorblockN{Willem Visser}
\IEEEauthorblockA{Stellenbosch University, South Africa\\
wvisser@cs.sun.ac.za 
}
}
\maketitle

\begin{abstract}
We introduce a new technique for inferring program invariants that uses symbolic
states generated by symbolic execution.  Symbolic states,
which consist of path conditions and constraints on local variables, 
are a compact description of sets of concrete program states and they
can be used for both invariant inference and invariant verification. 
Our technique uses a
counterexample-based algorithm that creates
concrete states from symbolic states, infers candidate
invariants from concrete states, and then verifies or refutes candidate
invariants using symbolic states.  The refutation case produces concrete
counterexamples that prevent spurious results and allow the technique to
obtain more precise invariants.  This process stops when the algorithm reaches
a stable set of invariants.

We present {\tool}, a tool that implements these ideas to automatically
generate invariants at arbitrary locations in a Java program.  The tool
obtains symbolic states from Symbolic PathFinder and uses existing algorithms to
infer complex (potentially nonlinear) numerical invariants. Our preliminary
results show that {\tool} is effective in using symbolic states to generate
precise and useful invariants for proving program safety and analyzing program
runtime complexity.  
We also show that {\tool} outperforms existing invariant generation systems.
\end{abstract}

\IEEEpeerreviewmaketitle
\input{intro}

\input{overview}
\input{invariants}
\input{syminfer}
\input{evaluation}

\input{related}
\input{conclusion}

\section*{Acknowledgment}
This material is based in part upon work supported by the
National Science Foundation under Grant Number 1617916.
\bibliographystyle{IEEEtran}
\balance
\bibliography{paper}

\end{document}

%% file: intro.tex
\section{Introduction}\label{sec:intro}

Program invariants describe properties that always hold at a program location.
Examples of invariants include pre/postconditions, loop invariants, and
assertions.
Invariants are useful in many programming tasks, including documentation, testing, debugging, verification, code generation, and synthesis~\cite{daikonernst,ernst2000dynamically,1629585,weimer06}.

Daikon~\cite{ernst2000dynamically} demonstrated that dynamic analysis is a practical approach to infer invariants from \emph{concrete program states} that are observed when running the program on sample inputs.
Dynamic inference is typically efficient and supports expressive invariants, but can often produce spurious invariants that do not hold for all possible inputs.
Several invariant generation aproaches (e.g., iDiscovery~\cite{idiscovery}, PIE~\cite{Padhi:2016:DPI:2908080.2908099}, ICE~\cite{Garg:2016:LIU:2837614.2837664}, NumInv~\cite{fse17}) use a hybrid approach that dynamically infers candidate invariants and then statically checks that they hold for all inputs.
For a spurious invariant, the checker produces counterexamples, which help the inference process avoid this invariant and obtain more accurate results.
This approach, called \emph{CounterExample Guided Invariant Generation} (CEGIR), iterates the inference and checking processes until achieving stable results.

In this paper, we present a CEGIR technique that uses \emph{symbolic program states}.
Our key insight is that symbolic states generated by a symbolic execution engine are
(a) compact encodings of large (potentially infinite) sets of concrete states, 
(b) naturally diverse since they arise along different execution paths, 
(c) explicit in encoding relationships between program variables, 
(d) amenable to direct manipulation and optimization, such as combining sets of states into a single joint encoding, and
(e) reusable across many different reasoning tasks within CEGIR algorithms.

We define algorithms for symbolic CEGIR that can be instantiated 
using different symbolic execution engines and present an
implementation {\tool} that uses Symbolic PathFinder~\cite{SPF} (SPF)---
a symbolic executor for Java.
{\tool} uses symbolic states in both the invariant inference and 
verification processes.
For inference, {\tool} uses symbolic states to generate 
concrete states to bootstrap a set of candidate invariants using
DIG~\cite{vuicse2012,vuicse2014,tosem2013}---which can infer expressive nonlinear invariants.
For verification, {\tool} formulates verification conditions from symbolic states
to confirm or refute an invariant, solves those using a SAT solver,
and produces counterexamples to refine the inference process.

Symbolic states allow {\tool} to overcome several limitations of 
existing CEGIR approaches.
iDiscovery, ICE, and PIE are limited to computing relatively simple invariants and  often do not consider complex programs with nonlinear arithmetic and properties such as $x=qy+r, x^2+y^2=z^2$.
These invariants appear in safety and security-critical software and can be leveraged to improve quality, e.g., to verify the absence of errors in Airbus avionic systems~\cite{CCF05} and to analyze program runtime complexity to detect security threats~\cite{hoffman17verifying, AGHKTW_PLDI2017}.
As our evaluation of {\tool} demonstrates in Sec.~\ref{sec:evaluation}, 
iDiscovery, which uses Daikon for inference, does not support nonlinear properties, and both ICE and PIE timeout frequently when nonlinear arithmetic is involved.
Recent work on NumInv~\cite{fse17} also
uses DIG to infer invariants, but it invokes KLEE~\cite{cadar2008klee} as a blackbox 
verifier for candidate invariants.  Since KLEE is unaware of the goals
of its verification it will attempt to explore the entire program state space
and must recompute that state space for each candidate invariant.
In contrast, {\tool} constructs a fragment of the state space
that generates a set of symbolic states that is sufficiently diverse
for invariant verification and it reuses symbolic states for all invariants.

We evaluated {\tool} over 3 distinct benchmarks which consist of 92 programs.
The study shows that {\tool}:
(1) can generate complex nonlinear invariants required in 21/27 NLA benchmarks, 
(2) is effective in finding nontrivial complexity bounds for 18/19 programs, with 4 of those improving on the best known bounds from the literature,
(3) improves on the state-of-the-art PIE tool in 41/46 programs in the HOLA benchmark, and 
(4) outperforms NumInv across the benchmarks while computing similar or better invariants.

These results strongly suggest that symbolic states form a powerful basis for computing program invariants.   They permit an approach that 
blends the best features of dynamic inference techniques and purely symbolic
techniques, such as weakest-precondition reasoning.
The key contribution of our work lies in the identification of the value
of symbolic states in CEGIR, in developing an algorithmic framework for
adaptively computing a sufficient set of symbolic states for invariant
inference, and in demonstrating, through our evaluation of 
{\tool}, that it improves on the best known techniques.

%% file: overview.tex
\section{Overview}\label{sec:overview}
\begin{figure}
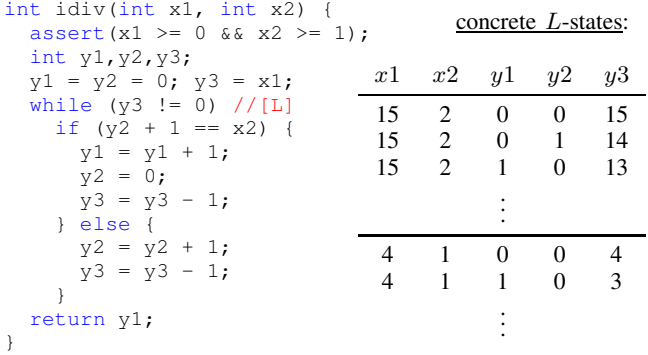

  \begin{minipage}{0.50\linewidth}
\begin{lstlisting}[numbers=none,mathescape,xleftmargin=0.2cm,emph={L}]
int idiv(int x1, int x2) {
  assert(x1 >= 0 && x2 >= 1);
  int y1,y2,y3;
  y1 = y2 = 0; y3 = x1;
  while (y3 != 0) //[L] 
    if (y2 + 1 == x2) {
      y1 = y1 + 1;
      y2 = 0;
      y3 = y3 - 1;
    } else {
      y2 = y2 + 1;
      y3 = y3 - 1;
    }
  return y1;
}
\end{lstlisting}
\end{minipage}
\hfill
\begin{minipage}{0.46\linewidth}
\centering
\small
\begin{tabular}{c  c c  c  c}
\multicolumn{5}{r}{\underline{concrete $L$-states}:}\\
\\
$x1$&$x2$&$y1$&$y2$&$y3$\\
\midrule
15& 2& 0& 0& 15\\
15& 2& 0& 1& 14\\
15& 2& 1& 0& 13\\
\multicolumn{2}{c}{}&\multicolumn{1}{c}{$\vdots$}&\multicolumn{1}{c}{}\\
\midrule            
4& 1& 0& 0& 4\\
4& 1& 1& 0& 3\\
\multicolumn{2}{c}{}&\multicolumn{1}{c}{$\vdots$}&\multicolumn{1}{c}{}\\
\end{tabular}
\end{minipage}
\caption{An integer division program and concrete $L$-states
  observed on inputs $(x1=15, ~x2=2)$ and $(x1=4,~x2=1)$}
\label{fig:example}
\end{figure}

We illustrate invariant inference using symbolic states on 
the integer division algorithm in
Figure~\ref{fig:example}; 
\texttt{L} marks the location at which we are interested
in computing invariants.  This example states assumptions on the
values of the parameters, e.g., no division by zero.
The best invariant at L is $\mathtt{x2} \cdot \mathtt{y1} + \mathtt{y2} + \mathtt{y3} =  \mathtt{x1}$.
This loop invariant encodes the precise semantics of 
the loop computing integer division, i.e.,
the dividend $\mathtt{x1}$ equals the divisor $\mathtt{x2}$ times the quotient $\mathtt{y1}$ plus the remainder, which is the sum of the two temporary variables $\mathtt{y2}$ and $\mathtt{y3}$.

Existing methods of dynamic invariant inference would 
instrument the program at location \texttt{L} to record values
of the 5 local variables, and then, given 
a set of input vectors, execute the program to record
a set of \textit{concrete states} of the program to generate
candidate invariants.  Since the focus here is on location 
\texttt{L}, we refer to these as \textit{L-states} and we
distinguish those that are \textit{observed} by instrumentation
on a program run.  It is these observed concrete L-states
that form the basis for all dynamic invariant inference techniques.

On eight hand-selected set of inputs that seek to expose diverse
concrete L-states, running Daikon~\cite{ernst2000dynamically} on this example results in very
simple invariants, e.g., $\mathtt{y1} \ge 0$, $\mathtt{x2} \ge 2$.
These are clearly much weaker than the desired invariant for this
example.  Moreover, the invariant on \texttt{x2} is actually
spurious since clearly $1$ can be passed as the second input
which will reach \texttt{L}.
Applying the more powerful DIG~\cite{tosem2013} invariant generator,
permits the identification of the desired invariant, but
it too will yield the spurious $\mathtt{x2} \ge 2$ invariant.
Spurious invariants are a consequence of the diversity
and representativeness of the inputs used, and the L-states
that are observed.  Leveraging symbolic states can help address
this weakness.

\input{ss-example}

\subsection{Generating a symbolic state space}
Figure~\ref{fig:symbolic-states} depicts a tree resulting from a depth-bounded
symbolic execution of the example.  The gray region includes paths
limited to at most 5 branches; in this setting depth is
a semantic property and syntactic branches with only a single
infeasible outcome are not counted, e.g., the branches with
labels enclosed in gray boxes.  We denote the unknown values of
program inputs using variables $X_i$ and return points with \texttt{ret}.

The states at location \texttt{L} are denoted $l_i$ in the figure.
An observed symbolic L-state, $l_i$, is defined by the conjunction 
of the path-condition,
i.e., the set of constraints on the tree-path to the state,
and $v_i$, a constraint that encodes the values of local variables
in scope at \texttt{L}.  For example, the symbolic state $l_2$ is defined as
$(X_2 = 1 \wedge X_1 \not= 0 \wedge X_1 \ge 0 \wedge X_2 \ge 1) \wedge (\mathtt{y1}=0 \wedge \mathtt{y2}=2 \wedge \mathtt{y3}=X_1-2)$.

As is typical in symbolic execution, it is possible to increase the
depth-bound and generate additional states, e.g., $l_6$, $l_{10}$,
$l_{12}$, and $l_{13}$ which all appear at a depth of 6 branches.

There are several properties of symbolic states that make them useful 
as a basis for efficient inference of invariants:

\paragraph{Symbolic states are expressive}
Dynamic analysis has to observe many concrete
L-states to obtain useful results.  Many of those states may be equivalent 
from a symbolic perspective.
A symbolic state, like $l_2$, encodes a potentially infinite set of
concrete states, e.g., $X_1 > 0 \wedge X_2 = 1$.  Invariant generation
algorithms can exploit this expressive power to account for 
the generation and refutation of candidate invariants from a
huge set of concrete states by processing a single symbolic state.

\paragraph{Symbolic states are relational}
Symbolic states encode the values of program variables as expressions
over free-variables capturing program inputs, i.e., $X_i$.  This
permits relationships between variables to be gleaned from the state.
For example, state $l_2$ represents the fact that $\mathtt{y3} < \mathtt{x1}$ for a large set of inputs.

\paragraph{Symbolic states can be reused}
Invariant generation has to infer or refute
candidate invariants relative to the set of observed concrete L-states.
This can grow in cost as the product of the number of candidates and
the size of number of observed states.
A disjunctive encoding of observed symbolic L-states, $\bigvee\limits_{i \in [1-13]} l_i$, can be constructed a single time and reused for each of
the candidate invariants, which can lead to performance improvement.

\paragraph{Symbolic states form a sufficiency test}
The diversity of symbolic L-states found during 
depth-bounded symbolic execution combined with the expressive power
of each of those states provides a rich basis for inferring strong
invariants.  We conjecture that for many programs a sufficiently
rich set of observed L-states for invariant inference will be
found at relatively shallow depth.   
For example, the invariants generated and not refuted by the
disjunction of L-states at depth 5, $L_{k=5} = \{l_1,l_2,l_3,l_4,l_5,l_7,l_8,l_9,l_{11}\}$, is the
same for those at depth 6, $\bigvee\limits_{i \in [1-13]} l_i$.
Consequently, we explore an adaptive and
incremental approach that increases depth only when new L-states
lead to changes in candidate invariants.  

\subsection{\tool\ \textrm{in action}}
{\tool} will invoke a symbolic executor to generate
a set of symbolic L-states at depth $k$, e.g., $k=5$ in our example for the gray region.
{\tool} then forms a small population of concrete L-states,
using symbolic L-states, to generate a set of candidate invariants
using DIG.
DIG produces three invariants at L for this example:
$\mathtt{y1} \cdot \mathtt{y2} \cdot \mathtt{y3} = 0$, 
$\mathtt{x2} \cdot \mathtt{y1} - \mathtt{x1} + \mathtt{y2} + \mathtt{y3} = 0$, and
$\mathtt{x1} \cdot \mathtt{y3} - 12 \cdot \mathtt{y1} \cdot \mathtt{y3} - \mathtt{y2} \cdot \mathtt{y3} - \mathtt{y3}^2  = 0$.
\tool\ attempts to refute these invariants by using the full
expressive power of the observed L-states to determine if 
all of the represented concrete states are consistent with the invariant.
It does this by calling a SAT solver to check implications such as
$\bigvee\limits_{l \in L_{k=5}} l \implies (\mathtt{y1} \cdot \mathtt{y2} \cdot \mathtt{y3} = 0)$.
This refutes the first and third candidate invariant.

\tool\ then seeks additional L-states by running symbolic execution
with a deeper bound, $k=6$.  While this process produces an additional 4
states to consider, none of those can refute the remaining
invariant candidate.  Thus, \tool\ terminates and produces the desired
invariant.


%% file: ss-example.tex
\begin{figure}[t]
\center
\pgfdeclarelayer{background}
\pgfsetlayers{background,main}
\begin{tikzpicture}[sloped, level distance=12mm, scale=0.95]
  \node[name=root] {$X_1 \ge 0 \wedge X_2 \ge 1$}
     child {
        node[circle, fill=black, xshift=0mm, yshift=0mm, inner sep=0mm] {\textcolor{white}{\small $l_1$}}
           child {
               node[circle, fill=black, xshift=2mm, yshift=0mm, inner sep=0mm] {\textcolor{white}{\small $l_2$}}
               child {
                   node[circle, fill=black, xshift=2mm, yshift=0mm, inner sep=0mm] {\textcolor{white}{\small $l_3$}}
                   child {
                      node[xshift=0mm, yshift=0mm] {}
                      child {
                         node[circle, fill=black, xshift=2mm, yshift=0mm, inner sep=0mm] {\textcolor{white}{\small $l_4$}}
                         child {
                            node[xshift=0mm, yshift=0mm] {}
                            child {
                               node[name=l5, circle, fill=black, xshift=2mm, yshift=0mm, inner sep=0mm] {\textcolor{white}{\small $l_5$}}
                               child {
                                  node[xshift=0mm, yshift=0mm] {}
                                  child {
                                     node[circle, fill=black, xshift=2mm, yshift=0mm, inner sep=0mm] {\textcolor{white}{\small $l_6$}}
                                     edge from parent node[above] {\tiny $X_1 \not= 4$}
                                  }
                                  child {
                                     node[xshift=-2mm, yshift=0mm] {\texttt{ret}}
                                     edge from parent node[above] {\tiny $X_1 = 4$}
                                  }
                                  edge from parent node[above] {\tiny \colorbox{gray!30}{$X_2 = 1$}}
                               }
                               edge from parent node[above] {\tiny $X_1 \not= 3$}
                            }
                            child {
                               node[name=l5ret, xshift=-2mm, yshift=0mm] {\texttt{ret}}
                               edge from parent node[above] {\tiny $X_1 = 3$}
                            }
                            edge from parent node[above] {\tiny \colorbox{gray!30}{$X_2 = 1$}}
                         }
                         edge from parent node[above] {\tiny $X_1 \not= 2$}
                      }
                      child {
                         node[xshift=-2mm, yshift=0mm] {\texttt{ret}}
                         edge from parent node[above] {\tiny $X_1 = 2$}
                      }
                      edge from parent node[above] {\tiny \colorbox{gray!30}{$X_2 = 1$}}
                   }
                   edge from parent node[above] {\tiny $X_1 \not= 1$}
               }
               child {
                   node[xshift=-2mm, yshift=0mm] {\texttt{ret}}
                   edge from parent node[above] {\tiny $X_1 = 1$}
               }
               edge from parent node[above] {\tiny $X_2 = 1$}
           }
           child {
               node[circle, fill=black, xshift=20mm, yshift=0mm, inner sep=0mm] {\textcolor{white}{\small $l_7$}}
               child {
                   node[xshift=0mm, yshift=0mm] {}
                   child {
                      node[circle, fill=black, xshift=-3mm, yshift=0mm, inner sep=0mm] {\textcolor{white}{\small $l_8$}}
                      child {
                         node[name=l9, circle, fill=black, xshift=2mm, yshift=0mm, inner sep=0mm] {\textcolor{white}{\small $l_9$}}
                         child {
                            node[xshift=0mm, yshift=0mm] {}
                            child {
                               node[name=l10, circle, fill=black, xshift=2mm, yshift=0mm, inner sep=0mm] {\textcolor{white}{\small $l_{10}$}}
                               edge from parent node[above] {\tiny $X_1 \not= 3$}
                            }
                            child {
                               node[xshift=0mm, yshift=0mm] {\texttt{ret}}
                               edge from parent node[above] {\tiny $X_1 = 3$}
                            }
                            edge from parent node[above] {\tiny \colorbox{gray!30}{$X_2 = 2$}}
                         }
                         edge from parent node[above] {\tiny $X_1 \not= 2$}
                      }
                      child {
                         node[xshift=-2mm, yshift=0mm] {\texttt{ret}}
                         edge from parent node[above] {\tiny $X_1 = 2$}
                      }
                      edge from parent node[above] {\tiny $X_2 = 2$}
                   }
                   child {
                      node[circle, fill=black, xshift=5mm, yshift=0mm, inner sep=0mm] {\textcolor{white}{\small $l_{11}$}}
                      child {
                         node[xshift=0mm, yshift=0mm] {}
                         child {
                            node[circle, fill=black, xshift=2mm, yshift=0mm, inner sep=0mm] {\textcolor{white}{\small $l_{12}$}}
                            edge from parent node[above] {\tiny $X_2 = 3$}
                         }
                         child {
                            node[circle, fill=black, xshift=-2mm, yshift=0mm, inner sep=0mm] {\textcolor{white}{\small $l_{13}$}}
                            edge from parent node[above] {\tiny $X_2 \not= 3$}
                         }
                         edge from parent node[above] {\tiny $X_1 \not= 2$}
                      }
                      child {
                         node[name=l11ret, xshift=-2mm, yshift=0mm] {\texttt{ret}}
                         edge from parent node[above] {\tiny $X_1 = 2$}
                      }
                      edge from parent node[above] {\tiny $X_2 \not= 2$}
                   }
                   edge from parent node[above] {\tiny $X_1 \not= 1$}
               }
               child {
                   node[xshift=-2mm, yshift=0mm] {\texttt{ret}}
                   edge from parent node[above] {\tiny $X_1 = 1$}
               }
               edge from parent node[above] {\tiny $X_2 \not= 1$}
           }
           edge from parent node[above] {\tiny $X_1 \not= 0$}
     }
     child {
        node[xshift=5mm, yshift=0mm] {\texttt{ret}}
        edge from parent node[above, xshift=1mm] {\tiny $X_1 = 0$}
     };

 \begin{pgfonlayer}{background}
    \filldraw[fill=gray!10] ($(root.north west)+(-2mm,0)$) -- ($(l5.south west)+(-4mm,-2mm)$) -- ($(l5ret.south east)+(0mm,-1mm)$) -- ($(l9.south west)+(0mm,-3mm)$) -- ($(l11ret.south east)+(4mm,-2mm)$) -- ($(root.north east)+(2mm,0)$) -- cycle;

    \node[name=ssl2, below of=l10, xshift=20mm] {\small $v_2 : \mathtt{y1}=1 \wedge \mathtt{y2}=0 \wedge \mathtt{y3} = X_1 - 1$};
    \node[name=ssl11, below of=ssl2] {\small $v_{11} : \mathtt{y1}=0 \wedge \mathtt{y2}=2 \wedge \mathtt{y3} = X_1 - 2$}; 
  \end{pgfonlayer}
\end{tikzpicture}
\caption{Symbolic Execution Tree and Symbolic L-states}
\label{fig:symbolic-states}
\end{figure}
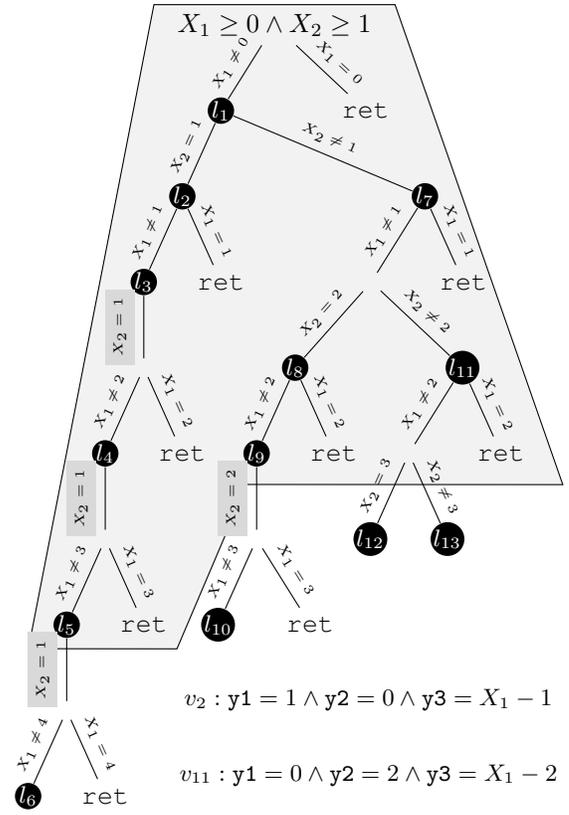

%% file: invariants.tex
\section{Dynamically Infer Numerical Invariants}\label{sec:dig}

\subsection{Numerical Invariants}
We consider invariants describing relationships over \emph{numerical} program variables such as $x \le y, 0 \le idx \le |arr| -1, x + 2y = 100$.
These numerical invariants have been used to verify program correctness, detect defects, establish security properties, synthesize programs, recover formal specifications, and more~\cite{ernst2000dynamically,slam,blast,ESP,z3ms,CCF05,leroy06,zeller2010,weimer06}.
A particularly useful class of numerical invariants involves \emph{nonlinear} relations, e.g., $x \le y^2, x2 \cdot y1 + y2 + y3 =  x1$.
While more complex these arise naturally in many safety-critical applications~\cite{CCF05,BCC03}.
\ignore{
They are required for implementations of common mathematical methods such as \texttt{mult, div, square, sqrt, mod} and have been used by the commercial tool Astree to verify the absense of errors in flight-control software~\cite{CCF05,BCC03}.
}

\begin{figure}[t]
\begin{lstlisting}[numbers=none,mathescape,xleftmargin=1.cm,emph={L}]
void pldi_fig2(int M, int N, int P){
  assert ($0 \le M$ && $0 \le N$ && $0 \le P$);
  int i = 0, j = 0, k = 0; 
  int t = 0; //counter variable
  while(i < N){//loop 1
    j = 0; t++;
    while(j < M){//loop 2	       
      j++; k = i; t++;
      while (k < P){// loop 3	
        k++; t++;
      }	    
      i = k;
    }
    i++;
  }
  // [L]
}
\end{lstlisting}
\caption{A program that has several nonlinear complexity bounds.}
\label{fig:complexity}
\end{figure}

In addition to capturing program semantics (e.g., as shown in Section~\ref{sec:overview}), nonlinear invariants can characterize the computational complexity of a program.
Figure~\ref{fig:complexity} shows a program, adapted from Figure 2 of~\cite{pldi09}, with 
nontrivial runtime complexity.
At first, this program appears to take $O(NMP)$ due to the three nested loops.
But closer analysis shows a more precise bound $O(N+NM+P)$ because the innermost loop 3, which is updated each time loop 2 executes, changes the behavior of the outer loop 1.

When analyzing this program, {\tool} discovers a complex nonlinear invariant over the variables $P,M,N$ and $t$ (a temporary variable used to count the number of loop iterations) at location $L$ (program exit):
\begin{equation*}
\small
\begin{aligned}
P^2Mt + PM^2t - PMNt - M^2Nt - PMt^2 + MNt^2 + PMt \\
- PNt - 2MNt + Pt^2 + Mt^2 + Nt^2 - t^3 - Nt + t^2 = 0.
\end{aligned}
\end{equation*}
This nonlinear (degree 4) equality looks very different than the expected bound $N+NM+P$ or even $NMP$.
However, when solving this equation (finding the roots of $t$), we obtain three solutions that describe the exact bounds of this program:
\begin{equation*}
\begin{aligned}
&t = 0               &\text{when}\quad& N =   0,\\
&t = P + M + 1       &\text{when}\quad& N \le P,\\
&t = N-M(P - N)      &\text{when}\quad& N >   P.
\end{aligned}
\end{equation*}
\noindent These results give more precise bounds than the given bound $N+MN+P$ in~\cite{pldi09}.

\subsection{Inferring Invariants using Concrete States}
To infer numerical invariants, {\tool} uses the algorithms in DIG~\cite{tosem2013}.
\ignore{
DIG is a dynamic analysis tool that infers invariants from a finite set of observed concrete states.
}
For numerical invariants, DIG finds (potentially nonlinear) equalities and inequalities.
Like other dynamic analysis tools, DIG generates \emph{candidate} invariants that only hold over observed concrete L-states.

\subsubsection{Nonlinear Equalities}\label{sec:dignle}

\begin{algorithm}[t]
\small
\DontPrintSemicolon
\SetKwInOut{Input}{input}
\SetKwInOut{Output}{output}

\SetKwFunction{extractEqts}{extractEqts}
\SetKwFunction{createTerms}{createTerms}
\SetKwFunction{createTemplate}{createTemplate}
\SetKwFunction{instantiate}{instantiate}
\SetKwFunction{solve}{solve}
\SetKwData{states}{states}
\SetKwData{numStates}{numStates}
\SetKwData{eqts}{eqts}
\SetKwData{template}{template}
\SetKwData{vars}{vars}
\SetKwData{terms}{terms}
\SetKwData{invs}{invs}
\SetKwData{inps}{inps}
\SetKwData{sols}{sols}
\SetKwData{eqInvs}{eqInvs}

\Input{\terms, \states}
\Output{equalities among \terms}
\BlankLine
\eqInvs $\leftarrow \emptyset$\;
$\template \leftarrow \createTemplate(\terms)$\;\label{line:template}
$\eqts \leftarrow \eqts \cup \instantiate(\template, \states)$\;\label{line:instantiate}
$\sols \leftarrow \solve(\eqts)$\;\label{line:eqtsolve}
$\eqInvs = \extractEqts(\sols, \terms)$\;\label{line:combine}
\KwRet \eqInvs
\caption{\texttt{inferEqts}: DIG's algorithm for finding candidate equalities}
\label{fig:inferEqts}
\end{algorithm}

To generate nonlinear equality invariants, DIG uses \emph{terms} to represent nonlinear information from the given variables up to a certain degree.
For example, the set of 10 terms $\{1,x,y,z,xy,xz,yz,x^2,y^2,z^2\}$ consist of all monomials up to degree $2$ over the variables $\{x,y,z\}$.

DIG then applies the steps shown in Figure~\ref{fig:inferEqts} to generate equality invariants over these terms using  concrete states observed at location L, and returns a set of possible equality relations among those terms.
First, we use the input terms to form an equation \emph{template} $c_1t_1 + c_2t_2 \dotsb + c_nt_n = 0$, where and $t_i$ are terms and $c_i$ are real-valued unknowns to be solved for (line~\ref{line:template}).
Next, we instantiate the template with concrete states to obtain concrete equations (line~\ref{line:instantiate}).
Then we use a standard equation solver to solve these equations for the unknowns (line~\ref{line:eqtsolve}).
Finally we combine solutions for the unknowns (if found) with the template to obtain equality relations (line~\ref{line:combine}).

\subsubsection{\it Octagonal Inequalities}
DIG uses various algorithms to infer different forms of inequality relations.
We consider the \emph{octagonal} relations of the form $c_1v_1 + c_2v_2\le k$ where $v_1,v_2$ are variables and $c_i\in \{-1,0,1\}$ and $k$ is real-valued.
These relations represent linear inequalities among program variables, e.g., $x \le y, -10 \le x - y \le 20$.
\ignore{
Octagonal invariants are useful in practice for checking array bound errors and memory leaks~\cite{mine2004weakly} and have been used for detecting bugs in avionic systems~\cite{CCF05,BCC03}.
}

To infer octagonal invariants from concrete states $\{(x_1,y_1), \dots\}$, we compute the upper and lowerbounds:
\begin{align*}
u_1=\max(x_i),&\; l_1=\min(x_i),\\\nonumber
u_2=\max(y_i),&\; l_2=\min(y_i),\\\nonumber
u_3=\max(x_i-y_i),&\; l_3=\min(x_i-y_i),\\\nonumber
u_4=\max(x_i+y_i),&\; l_4=\min(x_i+y_i)
\end{align*}
and form a set of 8 (octagonal) relations  $\{u_1 \ge x \ge l_1, u_2 \ge y \ge l_2,u_3 \ge x-y \ge l_3, u_4 \ge x+y \ge l_4\}$.

Although computing octagonal inequalities is very efficient (linear in the number of concrete states), the candidate results are likely spurious because the upper and lower bound values might not be in the observed concrete states.
{\tool} deals with such spurious invariants using a CEGIR approach described in Section~\ref{sec:cegir}.

%% file: syminfer.tex
\section{CEGIR Algorithms using Symbolic States}\label{sec:cegir}
\ignore{
{\tool} takes as input a program and a set locations, and returns invariants at those locations.
The key idea is using symbolic states to generate and verify program invariants.

\subsection{Concrete States and Symbolic States}
}

The behavior of a program at a location can be precisely represented by the set of all possible values of the variables in scope of that location.
We refer to these values as the \emph{concrete states} of the program.
Figure~\ref{fig:example} shows several concrete states observed at location $L$ when running the program on inputs  $(x1=15, x2=2)$ and $(x1=4,x2=1)$.

The set of all concrete states is the most precise representation of the relationship between variables at a program location, but it is potentially infinite and thus is difficult to use or analyze.
In contrast, invariants capture program behaviors in a much more compact way.
For the program in Figure~\ref{fig:example} invariants at location $L$ include: $0 \le x1, 1 \le x2, 0 \le y2 + y3, x2 \cdot y1 + y2 + y3 = x1, \dots$
The most useful at $L$ is $x2\cdot y1 + y2 + y3  = x1$, which describes the semantics of integer division.
The inequality $0 \le y2 + y3$ is also useful because it asserts that the remainder is non-negative.

Dynamic invariant generation techniques, like Daikon and DIG, use concrete program
states as inputs to compute useful invariants.
We propose to compute invariants from the \emph{symbolic states} of a program.
Conceptually, symbolic states serve as an intermediary representation
between a set of concrete program states and 
an invariant that might be inferred from those concrete states.

We assume a fixed and known set of variables in scope at
a given location in a program.  Moreover, we assume variables
are indexed and that for an index $i$, $var(i)$ is a
canonical name for that variable. 
Invariants will be inferred over these named variables.
This is straightforward for locals and parameters, but permits richer 
naming schemes for other memory locations.

We write a set of appropriately typed
values for those variables
as $\vec{v} \equiv \langle v_1, v_2, \ldots, v_n \rangle$,
where the indexing corresponds to that of variables.
Undefined variables 
have a $\bot$ value and the $i$th value is written $\vec{v}[i]$.
A \textit{concrete state} is 
$(l, \vec{v})$ where control is
at location $l$ and program variables have the values given by $\vec{v}$.

Let $I$ be a set free-variables that denote the undefined input
values of a program.
A \textit{symbolic value} is an expression written using constants,
elements of $I$, and the operators available for the value's type.
We write a sequence of symbolic values as
$\vec{e} \equiv \langle e_1, e_2, \ldots, e_n \rangle$.

\begin{definition}
A \textrm{symbolic state} is 
$(l, \vec{e}, c)$ where control is at location $l$, $c$ is
a logical formula written over $I$, and
a program variable takes on the corresponding concrete values
that are consistent with $c$ and symbolic value.
The semantics of a symbolic state is:
\[
\llbracket (l, \vec{e}, c)  \rrbracket =
\{ (l, \vec{v}) \mid SAT( (\bigwedge\limits_{i} \vec{v}[i]=\vec{e}[i]) \wedge c) \}
\]
\end{definition}
The role of $c$ in a symbolic state is to define the constraints
between variables, for example, that may be established on execution
paths reaching $l$---a path condition.

\subsection{Using Symbolic States}
Symbolic states can help invariant generation in many ways.
We describe two concrete techniques using symbolic states to generate diverse concrete states and to verify candidate invariants.
\subsubsection{Bootstrapping DIG with Concrete States}
\begin{algorithm}[t]
\small
\DontPrintSemicolon
\SetKwInOut{Input}{input}
\SetKwInOut{Output}{output}

\SetKwFunction{SAT}{SAT}
\SetKwFunction{getModel}{getModel}
\SetKwFunction{getStatesAt}{getStatesAt}
\SetKwFunction{eval}{eval}
\SetKwFunction{choose}{choose}

\SetKwData{symex}{symex}

\SetKwData{pcs}{pcs}
\SetKwData{block}{block}
\SetKwData{cstates}{cstates}
\SetKwData{sstates}{sstates}
\SetKwData{traces}{traces}

\Input{prog $P$, $L$, number of states $n$, depth $d$}
\Output{\cstates}
\BlankLine

$\block \gets false$\;
$\cstates \gets \emptyset$\;
$\sstates \gets \symex.\getStatesAt(P,L,d)$\;\label{line:getStatesAt}

\ForEach{$s \in \sstates$} {\label{line:loop1Start}
  \If{$SAT(s.c)$} {
    $\vec{i} \gets \getModel()$\;\label{line:model}
    $\cstates \gets \cstates \cup (L, \eval(s.\vec{e}, \vec{i}))$\;\label{line:union}
    $\block \gets \block \vee (\bigwedge\limits_{i \in I} \vec{i}[i]=i)$\;
  }
}

\While{$|\cstates| < n$}{\label{line:loop2Start}
  $s \gets \choose(\sstates)$\;
  \If{$SAT(s.c \wedge \neg \block)$} {
    $\vec{i} \gets \getModel()$\;
    $\cstates \gets \cstates \cup (L, \eval(s.\vec{e}, \vec{i}))$\;
    $\block \gets \block \vee (\bigwedge\limits_{i \in I} \vec{i}[i]=i)$\;
  }
}
\BlankLine
\KwRet \cstates
\caption{\texttt{genStates}: generate concrete states from symbolic states}
\label{fig:genStates}
\end{algorithm}

Our method generates candidate invariants using existing state of the
art concrete state-based invariant inference techniques like DIG.  
In this application we need only use a small number of concrete states
to bootstrap the algorithms to generate a diverse set of candidate
invariants since symbolic states will be used to refute spurious invariants.
In prior work~\cite{vuicse2012,tosem2013}, fuzzing was used to generate inputs and
that could be used here as well, but we can also exploit symbolic
states.

Figure~\ref{fig:genStates} shows how we use symbolic states to 
generate a diverse set of concrete states---at least one for each
symbolic state.
It first generates the set of symbolic L-states reachable depth
less than or equal to $d$ (line~\ref{line:getStatesAt}); note that these states can be cached
and reused for a given $P$ and $L$.

The loop on line~\ref{line:loop1Start} considers each such state, checks the satisfiability of the
states path condition, $c$, and then extracts the model from the solver.
We encode the model as a sequence, $\vec{i}$, indexed by the name of
a free input variables.  The symbolic state is then evaluated by
the binding of concrete values to input variables in the model.
This produces a concrete state which is accumulated.  
A conjunction of constraints equating the values of the model, 
$\vec{i}$, and the names of inputs, $I$, is added to the
blocking clause for future state generation.

The loop on line~\ref{line:loop2Start} generates additional concrete states up to
the requested number, $n$.  This process will randomly $\choose$
a symbolic state and then call the SAT solver to generate a
solution that has not already been computed; here $\vec{i}$ 
is converted to a conjunction of equality constraints between
input variables and values from a model.
When a solution is found, we use the same processing as in lines~\ref{line:model}-\ref{line:union}
to create a new concrete state.

\subsubsection{Symbolic States as a ``Verifier''}

\begin{algorithm}[t]
\small
\DontPrintSemicolon
\SetKwInOut{Input}{input}
\SetKwInOut{Output}{output}

\SetKwFunction{break}{break}
\SetKwFunction{SAT}{SAT}
\SetKwFunction{getModel}{getModel}
\SetKwFunction{getSymTraces}{getSymTraces}

\SetKwData{cex}{cex}
\SetKwData{False}{false}
\SetKwData{True}{true}
\SetKwData{unknown}{unknown}
\SetKwData{isInv}{isInv}
\SetKwData{symex}{symex}
\SetKwData{sstates}{sstates}
\SetKwData{vc}{vc}
\SetKwData{block}{block}
\SetKwData{result}{result}
\SetKwData{sat}{sat}
\SetKwData{unsat}{unsat}

\Input{prog $P$, loc $L$, prop $p$, clauses to \block}
\Output{counterexample \cex}
\BlankLine

$p.\isInv \leftarrow \unknown$\;
$\result \leftarrow \unknown$\;
$\result' \leftarrow \unknown$\;
$\cex \leftarrow \emptyset$\;
$k \leftarrow 10$ //default depth

\While{\True}{
  $\sstates \leftarrow \symex.\getStatesAt(P,L,k)$\;\label{line:extract}
  $\vc \leftarrow (\bigvee\limits_{s \in \sstates} (s.c \wedge \bigwedge\limits_{i} var(i)=s.\vec{e}[i])$\;\label{line:first}
  $\vc \gets \vc \wedge \neg (\bigvee \block))$\;\label{line:second}
  $\result' \leftarrow \SAT(\neg (\vc \implies p)$\;\label{line:check}
  \If{$\result' \equiv \result$}{
    $k \leftarrow k -1$\;
    \break
  }
  $\result \gets \result'$\label{line:check2}

  \If{$\result' \equiv \sat$}{\label{line:sat1}
    $p.\isInv \leftarrow \False$\;
    $\cex \leftarrow \getModel()$\; 
    \break\label{line:sat2}
  }\ElseIf{$\result' \equiv \unsat$}{
    $p.\isInv \leftarrow \True$    
  }\ElseIf{$\result' \equiv \unknown$}{
    $p.\isInv \leftarrow \unknown$
  }
  $k \leftarrow k + 1$\label{line:endloop}
}
\BlankLine
\KwRet \cex \;
\caption{\texttt{verify}: check a candidate property using symbolic states}
\label{fig:verify}
\end{algorithm}

Figure~\ref{fig:verify} shows how symbolic states are used to verify, or refute,
a property.
The algorithm obtains new symbolic states when it is determined that they
increase the accuracy of the verification.

Symbolic states are obtained from a symbolic execution engine.
There are potentially an infinite number of symbolic states at a location, 
but most existing symbolic execution tools have the ability to perform
a depth-limited search.
We wrap the symbolic execution engine to just return the symbolic L-states
encountered during search of a given depth (\getStatesAt).

The number of symbolic states varies with depth.
A low depth means few states.  Few states will tend to encode a small
set of concrete L-states, which limits verification and refutation
power.  Few states will also tend to produce a smaller and faster to solve
verification condition.  To address this cost-effectiveness tradeoff,
rather than try to choose an optimal depth, our algorithm
computes the lowest depth that yields symbolic states that change verification outcomes.
In essence, the algorithm
adaptively computes a good cost-effectiveness tradeoff for a given
program, location of interest, and invariant.

The algorithm iterates with each iteration considering a different depth, $k$. 
The body of the each iteration (lines~\ref{line:extract}\,--\,\ref{line:endloop}) works as follows.
It extract a set of symbolic states for the current depth using
symbolic execution (line~\ref{line:extract}); note
this can be done incrementally to avoid re-exploring the program's state space
using techniques like~\cite{Yang:2012:MSE:2338965.2336771}. 
It then formulates a verification condition out of three components.
(1) For each symbolic state,
it constructs the conjunction of its path condition, $c$, with constraints
encoding equality constraints between variables and their symbolic values, $\vec{e}$;
these per-state conjunctions are then disjoined.
This expresses the set of
concrete L-states corresponding to all of the symbolic states.
(2) The negation of the disjunction of the set of states that 
are to be blocked is formed.
These components are conjoined, which serves to eliminate the concrete
L-states that are to be blocked.
(3) If the resulting formula implies a candidate $p$ then that candidate
is consistent with the set of symbolic states.
We use a SAT solver to check the negation of this implication.

The solver can return \texttt{sat} which indicates that the property is 
not an invariant (lines~\ref{line:sat1}\,--\,\ref{line:sat2}).
The solver is also queried for a model which is a sample
state that is inconsistent with the proposed invariant.  This counterexample
state is saved so that the inference algorithm can search for
invariants that are consistent with it.
The solver can also return \texttt{unsat} indicating the property is a true invariant;
at least as far as the algorithm can determine given the symbolic states at the
current depth.
Finally, the solver can also return unknown, indicating it cannot determine whether the given property is true or false.

For the latter two cases, we increment the depth and explore a larger set
of symbolic states generated from a deeper symbolic execution.
Lines~\ref{line:check}\,--\,\ref{line:check2} work together to determine when increasing the depth 
does not influence the verification.  In essence, they check to see
whether the same result is computed at adjacent depths and if so, they
revert to the shallower depth and return. 

\subsection{A CEGIR approach using symbolic states}
\emph{CounterExample Guided Invariant Generation} (CEGIR) techniques consist of a guessing component that infers candidate invariants
and a checking component that verifies the candidate solutions.
If the candidate is invalid, the checker produces \emph{counterexamples}, i.e., concrete states that are not consistent with the candidate invariant.
The guessing process incorporates the generated counterexamples so that any
new invariants account for them.
Alternation of guessing and checking repeats until no candidates can be disproved.

{\tool} integrates symbolic traces into two CEGIR algorithms to compute candidate invariants.
These algorithms use the inference techniques described in Section~\ref{sec:dig} for equality and inequality invariants.

\subsubsection{Nonlinear Equalities}

\begin{algorithm}[t]
\small
\DontPrintSemicolon
\SetKwInOut{Input}{input}
\SetKwInOut{Output}{output}

\SetKwFunction{verify}{verify}
\SetKwFunction{infer}{infer}
\SetKwFunction{exec}{exec}
\SetKwFunction{break}{break}
\SetKwFunction{extractVars}{extractVars}
\SetKwFunction{createTerms}{createTerms}
\SetKwFunction{createTemplate}{createTemplate}
\SetKwFunction{instantiate}{instantiate}
\SetKwFunction{solve}{solve}
\SetKwFunction{inferEqts}{inferEqts}
\SetKwFunction{genStates}{genStates}

\SetKwData{states}{states}
\SetKwData{numStates}{numStates}
\SetKwData{NotEnoughTraces}{NotEnoughTraces}
\SetKwData{eqts}{eqts}
\SetKwData{template}{template}
\SetKwData{vars}{vars}
\SetKwData{terms}{terms}
\SetKwData{invs}{invs}
\SetKwData{cexs}{cexs}
\SetKwData{newcexs}{newcexs}
\SetKwData{block}{block}
\SetKwData{newcandidates}{newcandidates}
\SetKwData{candidates}{candidates}
\SetKwData{False}{False}
\SetKwData{True}{True}
\SetKwData{eqts}{eqts}
\SetKwData{sols}{sols}
\SetKwData{eqs}{eqs}
\SetKwData{sstates}{sstates}
\SetKwData{eqtInvs}{eqtInvs}
\SetKwData{isInv}{isInv}

\Input{program $P$, location \textsf{L}, degree $d$}
\Output{nonlinear equalities up to deg $d$ at \textsf{L}}
\BlankLine

$\states \leftarrow \emptyset$\;
$\invs \leftarrow \emptyset$\;
$\block \leftarrow \emptyset$\;
$\vars \leftarrow \extractVars(P, \textsf{L})$\;
$\terms \leftarrow \createTerms(\vars, d)$\;\label{line:terms}
$\states \leftarrow \genStates(P, L, |terms|)$\;\label{line:genstates}
$\candidates \leftarrow \inferEqts(\terms, \states)$\;\label{line:infer0}
\While{$\candidates \neq \emptyset$}{\label{line:loop0}
  $\cexs \leftarrow \emptyset$\;
  \ForEach {$p \in$ \candidates}{
    $\newcexs \leftarrow \verify(P, L, p, \block)$\;
    $\cexs \leftarrow \cexs \cup \newcexs$\;
    \lIf{$p.\isInv$}{$\invs \leftarrow \invs \cup \{p\}$}
  }

  \lIf{\cexs $\equiv \emptyset$}{\break}
  \block $\leftarrow \block \cup \cexs$\;
  \states $\leftarrow \states \cup \cexs$\;\label{line:append}
  $\newcandidates \leftarrow \inferEqts(\terms \states)$\;
  \candidates $\leftarrow \newcandidates - \invs$\;\label{line:loop1}
}

\BlankLine
\KwRet \invs \;

\caption{CEGIR algorithm for finding equalities.}
\label{fig:cegarEqt}
\end{algorithm}

Figure~\ref{fig:cegarEqt} defines our CEGIR algorithm
for computing non-linear equality invariants.
It consists of two phases: an initial invariant candidate generation
phase and then an iterative invariant refutation and refinment phase.

Lines~\ref{line:terms}\,--\,\ref{line:infer0} define the initial generation phase.
As as described in Section~\ref{sec:dignle}, we first create terms to represent nonlinear polynomials (line~\ref{line:terms}).
Because solving for $n$ unknowns requires at least $n$ unique equations, we need to  generate a sufficient set of concrete L-states (line~\ref{line:genstates}).
This can either be realized through fuzzing an instrumented
version of the program that records concrete L-states or,
as described in Figure~\ref{fig:genStates}, one can use symbolic
L-states to generate them.

The initial candidate set of invariants is iteratively refined
on lines~\ref{line:loop0}\,--\,\ref{line:loop1}.  The algorithm then refutes or confirms them using symbolic states as described in
Figure~\ref{fig:verify}.
Any property that is proven to hold is recorded in $invs$ and 
counterexample states, $cexs$, are accumulated across the set of properties.
Generated counterexample states are also blocked from contributing to the verification process.

If no property generated counterexample states, then the
algorithm terminates returning the verified invariants.
The counterexamples are added to the
set of states that are used to infer new candidate
invariants; this ensures that new invariants will be consistent
with the counterexample states (line~\ref{line:append}).  These new results may include some
already proven invariants, so we remove those from the set
of candidates considered in the next round of refinement.


\subsubsection{Octagonal Inequalities}\label{sec:cegirOct}
Our next CEGIR algorithm uses a divide and conquer approach to compute octagonal inequalities.
Given a term $t$, and an interval range $[minV, maxV]$, we compute the smallest integral upperbound $k$ of $t$ by repeatedly dividing the interval into halves that could contain $k$.
The use of an interval range $[minV, maxV]$ allows us to exclude terms ranges are too large (or that do not exist).
For example, if we check $t > maxV$ and it holds then we will not compute the bound of $t$ (which is strictly larger than $maxV$).

We start by checking a guess that $t \le midV$, where $midV = \lceil \frac{maxV + minV}{2} \rceil$.
These checks are performed by formulating a verification condition
from symbolic states in a manner that is analogous to Figure~\ref{fig:cegarEqt}.
If this holds, then $k$ is at most $midV$ and we tighten the search to a new interval $[minV , midV]$.
Otherwise, we obtain counterexample with $t$ having some value $c$, where $c > midV$.
We then tighten the search to a new interval $[c,maxV]$.
In either case, we repeat the guess for $k$ using an interval that is half the size of the previous one.
The search stops when $minV$ and $maxV$ are the same or their difference is one (in which case we return the smaller value if $t$ is less than or equal both).

To find octagonal invariants over a set of variables, e.g., $\{x,y,z\}$, we apply this method to find upperbounds of the terms $\{x, -x, y, -y, \dots, y+z, -y-z\}$. 
Note that we obtain both lower and upperbound using the same algorithm because the upperbound for $t$ essentially lowerbound of $-t$ since all computations are reversed for $-t$.

{\tool} reuses the symbolic states from the inference of equalities to
formulate verification conditions for inequalities.  This is another
example of how reuse speeds up inference.

%% file: evaluation.tex
\section{Implementation and Evaluation}\label{sec:evaluation}

We implemented {\tool} in Python/SAGE~\cite{sage}.
The tool takes as input a Java program with marked target locations and generates invariants at those locations.
We use Symbolic PathFinder (SPF)~\cite{SPF} to extract symbolic states for Java programs and the Z3 SMT Solver~\cite{z3ms} to check and produce models representing counterexamples.
We also use Z3 to check and remove redundant invariants.

{\tool} currently supports equality and inequality relations over numerical variables.
For (nonlinear) equalities, {\tool} uses techniques from DIG to limit the number
of generated terms.
This allows us, for example, to infer equalities up to degree 5 for a program with 4 variables and up to degree 2 for program with 12 variables.
For octagonal invariants, we consider upper and lower bounds within the range $[-10,10]$; we rarely observe inequalities with large bounds.
{\tool} can either choose random values in a range, $[-300,300]$ by default,
for bootstrapping, or use the algorithm in Figure~\ref{fig:genStates}.
\ignore{
  Our experience shows that we do not need very large input values to generate precise invariants.
  We start SPF with the depth 10,  which seems to be a good default search depth for all our test programs.
  }
All these parameters can be changed by {\tool}'s user; we chose these values based on our experience.

\subsection{Research Questions}
To evaluate {\tool}, we consider three research questions:
\begin{enumerate}
\item Is {\tool} effective in generating nonlinear invariants describing complex program semantics and correctness?
\item Can {\tool} generate expressive invariants that capture program runtime complexity?
\item How does {\tool} perform relative to PIE, a state-of-the-art invariant generation technique?
\end{enumerate}

To investigate these questions, we used 3 benchmark suites consist of 92 Java programs (described in details in each section).
These programs come with known or documented invariants.
\ignore{
 (e.g., assertions, loop invariants, postconditions).
}
Our objective is to compare {\tool}'s inferred invariants against these documented results.
To compare invariants, we used Z3 to check if the inferred results imply the documented ones.
We use a script to run {\tool} 11 times on each program and report the median results.
The scripts automatically terminates a run exceeding 5 minutes.
The experiments reported here were performed on a 10-core Intel i7 CPU 3.0GHZ Linux system with 32 GB of RAM.

\subsection{Analyzing Program Correctness}~\label{NLA}
In this experiment,  we use the NLA testsuite~\cite{tosem2013} which consists of 27 programs implementing mathematical functions such as \texttt{intdiv, gcd, lcm, power}.
Although these programs are relatively small (under 50 LoCs) they contain nontrivial structures such as nested loops and nonlinear invariant properties.
To the best of our knowledge, NLA contains the largest number of programs containing nonlinear arithmetic.
These programs have also been used to evaluate other numerical invariant systems~\cite{1236086,tosem2013,sharma2013data}.

These NLA programs come with known program invariants at various program locations (e.g., mostly nonlinear equalities for loop invariants and postconditions).
For this experiment, we evaluate {\tool} by finding invariants at these locations and comparing them with known invariants.

\subsubsection*{Results}
\begin{table}[t]
\caption{Experimental results for 27 programs in the NLA testsuite. \checkmark indicates when {\tool} generates results sufficiently strong enough to prove known invariants.
}
\footnotesize
\centering
\begin{tabular}{llc|crr|c}
\textbf{Prog}  & \textbf{Desc}&\textbf{Locs}  & \textbf{V, T, D}     & \textbf{Invs} &\textbf{Time (s)} &\textbf{Correct}\\
  \midrule
  cohendiv  &  int div     & 2  & 6,3,2  & 10  &  21.05 &   \checkmark \\ 
  divbin    &  int div     & 2  & 5,3,2  & 11  &  58.97 &   \checkmark \\ 
  manna     &  int div     & 1  & 5,4,2  & 6   &  35.33 &   \checkmark \\  
  hard      &  int div     & 2  & 6,3,2  & 6   &  29.40 &   \checkmark \\ 
  sqrt      &  square root & 1  & 4,4,2  & 5   &  20.03 &   \checkmark \\ 
  dijkstra  &  square root & 2  & 5,7,3  & 16  &  93.01 &   \checkmark \\  
  freire1   &  square root & 1  & -      & -   &  -     &   -          \\ 
  freire2   &  cubic root  & 1  & -      & -   &  -     &   -          \\ 
  cohencu   &  cubic sum   & 1  & 5,5,3  & 4   &  21.90 &   \checkmark \\ 
  egcd1     &  gcd         & 1  & 8,3,2  & 14  &  122.22&   \checkmark \\ 
  egcd2     &  gcd         & 2  & -      & -   &  -     &   -          \\   
  egcd3     &  gcd         & 3  & -      & -   &  -     &   -          \\  
  prodbin   &  gcd, lcm    & 1  & 5,3,2  & 7   &  56.17 &   \checkmark \\ 
  prod4br   &  gcd, lcm    & 1  & 6,3,3  & 9   &  84.37 &   \checkmark \\ 
  knuth     &  product     & 1  & -      & -   &  -     &   -          \\  
  fermat1   &  product     & 3  & 5,6,2  & 17  &  60.26 &   \checkmark \\
  fermat2   &  divisor     & 1  & 5,6,2  & 8   &  36.83 &   \checkmark \\
  lcm1      &  divisor     & 3  & 6,3,2  & 24  &  248.17&   \checkmark \\ 
  lcm2      &  divisor     & 1  & 6,3,2  & 7   &  34.17 &   \checkmark \\ 
  geo1      &  geo series  & 1  & 4,4,2  & 8   &  158.27&   \checkmark \\ 
  geo2      &  geo series  & 1  & 4,4,2  & 9   &  147.75&   \checkmark \\ 
  geo3      &  geo series  & 1  & -      & -   &  -     &   -          \\   
  ps2       &  pow sum     & 1  & 3,3,2  & 3   &  18.39 &   \checkmark \\ 
  ps3       &  pow sum     & 1  & 3,4,3  & 3   &  19.69 &   \checkmark \\ 
  ps4       &  pow sum     & 1  & 3,4,4  & 3   &  19.92 &   \checkmark \\
  ps5       &  pow sum     & 1  & 3,5,5  & 3   &  46.19 &   \checkmark \\
  ps6       &  pow sum     & 1  & 3,5,6  & 3   &  41.19 &   \checkmark \\
\bottomrule
\end{tabular}
\label{table:nla}
\end{table}

Table~\ref{table:nla} shows the results of {\tool} for the 27 NLA programs.
Column \textbf{Locs} show the number of locations where we obtain invariants.
Column \textbf{V,T,D} shows the number of variables, terms, and highest degree from these invariants.
Column \textbf{Invs} shows the number of discovered equality and inequality invariants.
Column \textbf{Time} shows the total time in seconds.
Column \textbf{Correct} shows if the obtained results match or imply the known invariants.

For 21/27 programs, {\tool} generates correct invariants that match or imply the known results.
In most cases, the discovered invariants match the known ones exactly.
Occasionally, we obtain results that are equivalent or imply the known results.
For example, for \texttt{sqrt}, for some runs we obtained the documented equalities $t=2a+1, s=(a+1)^2$, and for other runs we obtain $t=2a+1, t^2 - 4s + 2t = -1$, which are equivalent to $s=(a+1)^2$ by replacing $t$ with $2a+1$.
We also obtain undocumented invariants, e.g., {\tool} generates the postconditions $x = qy+r, 0 \le r, r \le x, r \le y -1$ for \texttt{cohendiv}, which computes the integer division result of two integers $q = x \div y$, 
The first invariant is known and describes the precise semantics of integer division: the dividend $x$ is the divisor $y$ times the quotion $q$ plus the remainder $r$.
The other obtained inequalities were undocumented.
For example, $r \ge 0$ asserts that the remainder $r$ is non-negative and $r \le x, r\le y -1$ state that $r$ is at most the dividend $x$, but is strictly less than the divisor $y$.
Our experience shows that {\tool} is capable of generating many invariants that are unexpected yet correct and useful. 

{\tool} did not find invariants for 6/27 programs (marked with ``-'' in Table~\ref{table:nla}).
For \texttt{egcd2}, \texttt{egcd3}, the equation solver used in SAGE takes exceeding long time for more than half of the runs.
For \texttt{geo3}, we obtained the documented invariants and others, but Z3 stops responding when checking these results.
\texttt{freire1} and \texttt{freire2} contain floating point arithmetic, which are currently not supported by {\tool}.
SPF failed to produce symbolic states for \texttt{knuth} for any depth we tried.
This program invokes a library function \texttt{Math.sqrt} and SPF does not know the semantics of this function and thus fails to provide useful symbolic information.
For \texttt{egcd2}, \texttt{egcd3}, and \texttt{geo3}, {\tool} times out
after 5 minutes, and for \texttt{freire1}, \texttt{freire2}, and \texttt{knuth}, it exits upon encountering the unsupported feature.

\subsection{Analyzing Computational Complexity}

\begin{table}
\caption{Experimental results for computing programs' complexities. \checkmark: {\tool} generates the expected bounds.
\checkmark$^*$: program was slightly modified to assist the analysis. 
\checkmark \checkmark: {\tool} obtains more precise bounds than reported results.}
\footnotesize
\centering
\begin{tabular}{l|ccr|c}
\textbf{Prog}  & \textbf{V, T, D}     & \textbf{Invs} & \textbf{Time (s)} &\textbf{Bound}\\
  \midrule
  cav09\_fig1a	   & 2,5,2    &   1  & 12.41      & \checkmark            \\
  cav09\_fig1d     & 2,5,2    &   1  & 12.44      & \checkmark            \\
  cav09\_fig2d     & 3,2,2    &   3  & 58.40      & \checkmark            \\
  cav09\_fig3a     & 2,2,2    &   3  & 8.75       & \checkmark            \\
  cav09\_fig5b     & 3,5,2    &   6  & 49.44      & \checkmark$^*$        \\
  pldi09\_ex6      & 3,8,3    &   6  & 57.00      & \checkmark            \\ 
  pldi09\_fig2     & 3,15,4   &   6  & 60.60      & \checkmark\checkmark  \\
  pldi09\_fig4\_1  & 2,3,1    &   3  & 56.24      & \checkmark            \\
  pldi09\_fig4\_2  & 4,4,2    &   5  & 28.32      & \checkmark            \\
  pldi09\_fig4\_3  & 3,3,2    &   3  & 59.19      & \checkmark            \\ 
  pldi09\_fig4\_4  & 5,4,2    &   -  & -          & -                     \\
  pldi09\_fig4\_5  & 3,4,2    &   3  & 103.70     & \checkmark$^*$        \\
  popl09\_fig2\_1  & 5,12,3   &   2  & 50.86      & \checkmark\checkmark  \\
  popl09\_fig2\_2  & 4,9,3    &   2  & 53.48      & \checkmark\checkmark  \\
  popl09\_fig3\_4  & 3,4,3    &   4  & 58.62      & \checkmark            \\
  popl09\_fig4\_1  & 3,3,2    &   4  & 65.19      & \checkmark$^*$        \\
  popl09\_fig4\_2  & 5,12,3   &   2  & 51.24      & \checkmark\checkmark  \\
  popl09\_fig4\_3  & 3,3,2    &   5  & 31.57      & \checkmark            \\
  popl09\_fig4\_4  & 3,3,2    &   3  & 36.89      & \checkmark            \\
\bottomrule
\end{tabular}
\label{table:complexity}
\end{table}

As shown in Section~\ref{sec:dig}, nonlinear invariants can represent precise  program runtime complexity.
More specifically, we compute the roots of nonlinear relationships to obtain disjunctive information (e.g., $x^2 = 4 \implies (x=2 \vee x = -2)$, which capture different and precise complexity bounds of programs.

To further evaluate {\tool} on discovering program complexity, we collect 19 programs, adapted from existing static analysis techniques specifically designed to find runtime complexity~\cite{pldi09,Gulwani:2009:SSC:1575060.1575069,DBLP:conf/popl/GulwaniMC09}\footnote{We remove nondeterministic features in these programs because {\tool} assumes determinstic behaviors.}.
These programs, shown in Table~\ref{table:complexity}, are small, but contain nontrivial structures and represent examples from Microsoft's production code~\cite{pldi09}.
For this experiment, we instrument each program with a fresh variable $t$ representing the number of loop iterations and generate postconditions over $t$ and input variables (e.g., see Figure~\ref{fig:complexity}).

\subsubsection*{Results}
Table~\ref{table:complexity} shows the median results of {\tool} from 11 runs.
Column~\textbf{Bound} contains a \checkmark if we can generate invariants matching the bounds reported in the respective work, and \checkmark \checkmark if the discovered invariants represent more precise bounds than the reported ones.
A \checkmark$^*$ indicates when the program was modified slightly to help our analysis---described below.

For 18/19 programs, {\tool} discovered runtime complexity characterizations 
that match or improve on reported results.
For \texttt{cav09\_fig1a}, we found the invariant $mt - t^2 - 100m + 200t = 10000$, which indicates the correct bound $t = m + 100 \vee t = 100$.
For these complexity analyses, we also see the important role of combining both inequality and equality relations to produce informative bounds.
For \texttt{popl09\_fig3\_4}, {\tool} inferred nonlinear equality showing that $t=n\vee t=m$ and inequalities asserting that $t\ge n \wedge t \ge m$, together indicating that $t=\mathsf{max}(n,m)$, which is the correct bound for this program.
In four programs, {\tool} obtains better bounds than reported results.
The \texttt{pldi\_fig2} programs showing in Figure~\ref{fig:complexity} is a concrete example where the obtained three bounds are strictly less than the given bound.

For several programs we needed some manual instrumentation or inspections to help the analysis.
For \texttt{popl09\_fig4\_1} we added the precondition asserting the input $m$ is nonnegative.
For \texttt{pldi09\_fig4\_5}, we obtained nonlinear results giving three bounds $t=n-m$, $t=m$, and $t=0$, which establish the reported upperbound $t=max(0,n-m,m)$.
For \texttt{pldi09\_fig4\_4}, we obtained invariants that are insufficient to show the reported bound.
However, if we create a new term representing the quotient of an integer division of two other variables in the program, and obtain invariants over that term, we obtain more precise bounds than those reported.

\subsection{Comparing to PIE}
\begin{table}
\footnotesize
\caption{{\tool} run on HOLA benchmarks. \checkmark: produce sufficiently strong results to prove assertions. $\circ$: fail to make sufficiently strong invariants.}
\centering
\begin{tabular}{c | c | c | c}
  \textbf{Benchmark} & \textbf{PIE time (s)} & \textbf{{\tool} time (s)} & \textbf{Correct} \\
  \midrule
  H01 & 21.88 &    3.69 & $\circ$     \\ 
  H02 & 36.12 &    3.36 & \checkmark \\ 
  H03 & 56.28 &   23.96 & \checkmark \\
  H04 & 19.11 &    3.12 & \checkmark \\ 
  H05 & 25.19 &    3.76 & \checkmark \\ 
  H06 & 61.98 &    4.56 & \checkmark \\
  H07 & -     &    4.58 & \checkmark \\
  H08 & 19.02 &    4.33 & \checkmark \\  
  H09 & -     &   19.66 & \checkmark \\ 
  H10 & 24.6  &    4.25 & \checkmark \\
  H11 & 27.95 &    5.13 & \checkmark \\
  H12 & 44.52 &   14.60 & \checkmark \\
  H13 & -     &    3.99 & \checkmark \\
  H14 & 25.98 &    4.07 & \checkmark \\  
  H15 & 48.30 &    4.20 & \checkmark \\
  H16 & 33.19 &    4.99 & \checkmark \\
  H17 & 53.36 &    3.03 & \checkmark \\
  H18 & 21.70 &    5.69 & \checkmark \\
  H19 & -     &    5.05 & \checkmark \\
  H20 & 331.93&   29.45 & \checkmark\\
  H21 & 25.65 &   18.99 &  $\circ$ \\ 
  H22 & 25.40 &    4.50 & \checkmark \\
  H23 & 23.40 &    4.90 & \checkmark \\
  H24 & 51.22 &      -  & - \\ 
  H25 & -     &    4.31 & \checkmark\\
  H26 & 87.64 &    5.55 & \checkmark \\  
  H27 & 55.41 &     -   & - \\  
  H28 & 22.16 &    6.37 & \checkmark \\
  H29 & 58.82 &    6.80 & \checkmark \\
  H30 & 33.92 &    4.42 & $\circ$    \\  
  H31 & 88.10 &   38.94 & \checkmark \\
  H32 & 226.73&    6.75 & \checkmark \\
  H33 & -     &    6.95 & \checkmark \\
  H34 & 121.87&   11.34 & \checkmark \\
  H35 & 20.07 &    3.47 & \checkmark \\
  H36 & -     &    7.61 & \checkmark \\  
  H37 & -     &    9.87 & \checkmark \\
  H38 & 37.37 &    6.47 & \checkmark \\ 
  H39 & 24.68 &    3.99 & \checkmark \\
  H40 & 60.71 &   60.20 & \checkmark \\  
  H41 & 34.10 &    6.89 & \checkmark \\
  H42 & 54.93 &    5.55 & $\circ$    \\
  H43 & 21.16 &    5.34 & \checkmark \\
  H44 & 31.92 &   13.67 & \checkmark \\
  H45 & 84.00 &    5.39 & \checkmark \\
  H46 & 27.56 &    6.21 & \checkmark \\ 
  \bottomrule
\end{tabular}
\label{table:hola}
\end{table}

We compare {\tool} to the recent CEGIR-based invariant tool PIE~\cite{Padhi:2016:DPI:2908080.2908099}.
PIE aims to verify annotated relations by generating invariants based 
on the given assertions.
In contrast, {\tool} generates invariants at given locations without given assertions or postconditions.
We use the HOLA benchmarks~\cite{Dillig:2013:IIG:2509136.2509511}, adapted by the PIE developers.
These programs are annotated with various assertions representing loop invariants and postconditions.
This benchmark consists of 49 small programs, but contain nontrivial structures including nested loops or multiple sequential loops.
These programs,  shown in Table~\ref{table:hola}, have been used as benchmarks for other static analysis techniques~\cite{beyer2007software,Gupta:2009:IEI:1575060.1575112,jeannet2010interproc}.

For this experiment, we first run PIE and record its run time on proving the annotated assertions.
Next, we removed the assertions in the programs and asked {\tool} to generate invariants at those locations.
Our objective is to compare {\tool}'s discovered invariants with the annotated assertions.
Because these HOLA programs only consist of assertions having linear relations, we ask {\tool} to only generate invariants up to degree 2 (quadratic relations can represent linear relations, e.g., $x^2=4 \implies x=2 \vee x=-2$).

\subsubsection*{Results}
Table~\ref{table:hola} shows these obtained results from PIE and {\tool}.
Column \textbf{PIE time} shows the time, in seconds, for PIE to run each program.
Column \textbf{{\tool} time} shows the time, in seconds, for {\tool} to generate invariants for each program (the median of 11 runs).
The ``-'' symbol indicates when PIE fails to prove the given assertions, e.g., because it generates invariants that are too weak.
Column \textbf{Correct} shows whether {\tool}'s generated invariants match or imply the annotated assertions and therefore prove these assertions.
For this experiment we manually check the result invariants and use Z3 to compare them to the given assertions.
A \checkmark indicates that the generated invariants match or imply the assertions.
A $\circ$ indicates that the generated invariants are not sufficiently strong to prove the assertions.

For 40/46 programs, {\tool} discovered invariants are sufficiently strong to prove the assertions.
In most of these cases we obtained correct and stronger invariants than the given assertions.
For example, for H23, {\tool} inferred the invariants $i = n, n^2 - n - 2s = 0, -i \le n$, which imply the postcondition $s \ge 0$.
For H29, we obtained the invariants $b + 1 = c, a + 1 = d, a + b \le 2, 2 \le a$, which imply the given postcondition $a+c = b+d$.

Surprisingly, {\tool} also found invariants that are precise enough to establish conditions under forms that are \emph{not} supported by {\tool}.
For example, H8 contains a postcondition $x<4\ \vee\ y>2$, which has a disjunctive form of strict inequalities.
{\tool} did not produce this invariant, but instead produced a correct and stronger relation $x \le y$, which implies this condition.
Many HOLA programs contain disjunctive (or conditional) properties, e.g., \lt{if(c) assert (p);} where the property $p$ only holds when the condition $c$ holds (written $c\implies p$).
For example, for H18, we obtained $fj = 100f$, which implies the conditional assertion $f \ne 0 \implies j=100$.
For H37,  PIE failed to prove the postcondition \lt{if (n > 0) assert(0 <= m && m < n);} which involves both conditional assertions and strict inequalities.
For this program, {\tool} inferred 2 equations and 3 inequalities\footnote{$m^2 = nx - m - x, mn = x^2 - x,  -m  \le x, x \le m + 1, n \le x$}, which together establish the postcondition.

For 6/46 programs, {\tool} either failed to produce invariants (2 programs marked with ``-'') or discovered invariant that are not strong enough to prove the given assertions (4 programs marked with $\circ$).
For both H24 and H27, Z3 stops responding when checking the inferred results and the run were terminated after 5 minutes.
For H01, we found the invariant $x  = y$, which is not sufficiently to establish the postcondition $y\le 1$.
For H27, {\tool} found no relation involving the variable $c$ to prove the assertion $c \ge 0$.


\paragraph*{Summary} These preliminary results show {\tool} generates expressive, useful, and interesting invariants describing the semantics and match documented invariants (21/27 NLA programs), discovers difficult invariants capturing precise and informative complexity bounds of programs (18/19 programs), and is competitive with PIE (40/46 HOLA programs).
We also note that PIE, ICE, and iDiscovery (another CEGIR-based tools reviewed in Section~\ref{sec:related}), cannot find any of these high-degree nonlinear invariants found by {\tool}.

\subsection{Threats to Validity}
{\tool}'s run time is dominated by computing invariants, more specifically solving hundred of equations for hundred of unknowns.
The run time of DIG can be improved significantly by limiting the search to invariants of a given maximum degree rather than using the default setting. 
Verifying candidate invariants, i.e., checking implication using the Z3 solver, is much faster than DIG, even when multiple checks are performed at different depths. 
This shows an advantage of reusing symbolic states when checking new invariants.

{\tool} encodes all symbolic states to into the Z3 verification condition.
This results in complex formulas with large disjunctions that can make Z3 timeout.
Moreover, depending on the program, SPF might not be able to 
generate all possible symbolic states.
In such cases, {\tool} cannot refute candiate invariants and thus may produce \emph{unsound} results.
However, our experience shows that SPF, by its nature as a symbolic executor, turns out to be very effective in producing sufficient symbolic states, which effectively remove invalid candidates.


\ignore{We believe that the availability of the extracted symbolic states will permit optimization of these
conditions.  For example, they can be sliced relative to the candidate
invariant.  Since the combination of symbolic states is disjunctive, we
can also form a set of verification conditions, from a partition of the
states, and solve each in parallel.
}
\ignore{
The SAGE equation solver occasionally hangs when solving certain large sets of equations.
We are sending these examples to the SAGE developers, but it might be worth to explore other tool such as Matlab.
}

Finally, we reuse existing analysis tools, such as DIG and SPF, which provides
a degree of assurance in the correctness of {\tool}, but our primary
means of assuring internal validity was performing both manual and automated (SMT) 
checking of the invariants computed for all subject programs.
While our evaluation uses a variety of programs from different benchmarks,
these programs are small and thus do not represent large software projects.
Their use does promote comparative evaluation and reproducibility of our
results.
We believe using symbolic states will allow for the generation of
useful and complex invariants for larger software systems, in part because 
of the rapid advances in symbolic execution and SMT solving technologies
and {\tool} leverages those advances.

%% file: related.tex
\section{Related Work and Future Work}\label{sec:related}
Daikon~\cite{ernst2000dynamically} is a well-known dynamic tool that infers candidate invariants 
under various templates over concrete program states. 
The tool comes with a large set of templates which it tests against observed
concrete states, removing those that fail, and return the remaining ones as candidate invariants.
DIG~\cite{tosem2013} is similar to Daikon, but focuses on numerical invariants and therefore can compute more expressive numerical relations than those supported by Daikon's templates.

PIE~\cite{Padhi:2016:DPI:2908080.2908099} and ICE~\cite{Garg:2016:LIU:2837614.2837664} uses CEGIR to infer invariants to prove a given specification.
To prove a property, PIE iteratively infers and refined invariants by constructing necessary predicates to separate (good) states satisfying the property and (bad) states violating that property.
ICE uses a decision learning algorithm to guess inductive invariants over predicates separating good and bad states.
The checker produces good, bad, and ``implication'' counterexamples to help 
learn more precise invariants.
For efficiency, they focus on octagonal predicates and only search for invariants that are boolean combinations of octagonal relations.
In general, these techniques focus on invariants that are necessary to prove a given
specification and, thus, the quality of the invariants are dependent target specification.

NumInv~\cite{fse17} is a recent CEGIR tool that discovers invariants for C programs.
The tool also uses DIG's algorithms to infer equality and inequality relations.
For verification it instruments invariants into the program and runs the KLEE test-input generation tool~\cite{cadar2008klee}.
KLEE does use a symbolic state representation internally, but this is inaccessible to
NumInv.  Moreover, KLEE is unaware of its use in this context and it recomputes the
symbolic state space completely for each verification check, which is inefficient.
For the experiments in Section~\ref{sec:evaluation}, {\tool} is comparable to NumInv in
the quality of invariants produced, but {\tool} runs faster in spite of the
fact that KLEE's symbolic execution of C programs is known to be faster than SPF's
performance on Java programs.  We credit this to the benefits of using symbolic states.

Similar to {\tool}, the CEGIR-based iDiscovery~\cite{idiscovery} tool uses SPF to check invariants.
However, iDiscovery does not exploit the internal symbolic state representation of symbolic excution but instead runs SPF as a blackbox to check program assertions encoding candidate invariants.
To speed up symbolic execution, iDiscovery applies several optimizations such as using the Green solver~\cite{Visser:2012:GRR:2393596.2393665} to avoid recomputing the symbolic state space for each check.
In contrast, {\tool} precomputes the full disjunctive SMT formula encoding the paths to the interested location once and reuses that formula to check candidate invariants.
For dynamic inference, iDiscovery uses Daikon and thus has limited support for numerical invariants.
For example, iDiscovery cannot produce the required nonlinear invariants or any relevant inequalities for the programs in Figures~\ref{fig:example} and~\ref{fig:complexity}.
Note that for programs involving non-numerical variables, Daikon/iDiscovery might be able to infer more invariants than {\tool}.

{\tool} is unlike any of the above in its reliance on symbolic states to
bootstrap, verify and iteratively refine the invariant generation process.
There are clear opportunities for significantly improving the performance of {\tool}
and targeting different languages, such as C through the use of other symbolic
executors.   For example, generating symbolic states can be sped up for invariant inference
by combining directed symbolic execution~\cite{Ma:2011:DSE:2041552.2041563} to target locations
of interest, memoized symbolic execution~\cite{Yang:2012:MSE:2338965.2336771} to store symbolic
execution trees for future extension, and parallel symbolic execution~\cite{Staats:2010:PSE:1831708.1831732} to accelerate the incremental generation of the tree.
Moreover, we can apply techniques for manipulating symbolic states in symbolic execution~\cite{cadar2008klee,Visser:2012:GRR:2393596.2393665}
to significantly reduce the complexity of the verification conditions sent to the solver.

%% file: conclusion.tex
\section{Conclusion}\label{sec:conclusion}
We present {\tool} a method that uses symbolic encodings of program
states to efficiently discover rich invariants over numerical variables at 
arbitrary program locations.
{\tool} uses a CEGIR approach that uses symbolic states to generate candidate invariants and also to verify or refute, and iteratively
refine, those candidates.
Key to the success of {\tool} is its ability to directly manipulate
and reuse rich encodings of large sets of concrete program states.
Preliminary results on a set of 92 nontrivial programs show that {\tool} is effective in discovering useful invariants to describe precise program semantics, characterize the runtime complexity of programs, and verify nontrivial correctness properties.